\documentstyle[aps,prl,preprint,floats,epsfig]{revtex}  

\def\DESepsf(#1 width #2){\epsfxsize=#2 \epsfbox{#1}}
\begin{document}

\pagestyle{empty}    

\epsfysize3cm
\epsfbox{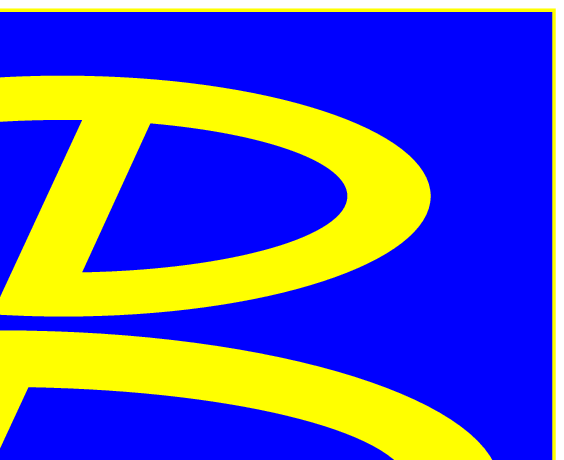}                                   

\begin{flushright}
\vskip -2.5cm
\noindent
\hspace*{4.in}
\begin{minipage}{1.7in}
KEK preprint 2001-117
 
\vspace*{-.3cm} 
Belle preprint 2001-14 
\end{minipage}
\end{flushright}
\vskip 1cm 

\begin{center}
{\bf \Large
Observation of Color-suppressed $\overline{B}{}^0 \to
D^{0}\pi^0$, $D^{*0}\pi^0$, $D^0\eta$, and $D^0\omega$
Decays\footnote{to appear in PRL} 
}
\end{center}






\tighten

\normalsize


\vskip2pc
\begin{center}
  K.~Abe$^{9}$,               
  K.~Abe$^{36}$,              
  R.~Abe$^{27}$,              
  I.~Adachi$^{9}$,            
  Byoung~Sup~Ahn$^{15}$,      
  H.~Aihara$^{38}$,           
  M.~Akatsu$^{20}$,           
  Y.~Asano$^{42}$,            
  T.~Aso$^{41}$,              
  V.~Aulchenko$^{2}$
% ...
  \vskip1pc {\bf \large Belle Collaboration}
\end{center}

\small
\begin{center}
% ...
$^{2}${Budker Institute of Nuclear Physics, Novosibirsk}\\
% ...
$^{9}${High Energy Accelerator Research Organization (KEK), Tsukuba}\\
% ...
$^{15}${Korea University, Seoul}\\
% ...
$^{20}${Nagoya University, Nagoya}\\
% ...
$^{27}${Niigata University, Niigata}\\
% ...
$^{36}${Tohoku Gakuin University, Tagajo}\\
% ...
$^{38}${University of Tokyo, Tokyo}\\
% ...
$^{41}${Toyama National College of Maritime Technology, Toyama}\\
$^{42}${University of Tsukuba, Tsukuba}\\
% ...
(\today)
\end{center}

\normalsize

\begin{abstract}
We report the first observation of color-suppressed
$\overline{B}{}^0 \to D^0 \pi^0$, $D^{*0} \pi^0$, $D^{0} \eta$, 
and $D^{0}\omega$ decays and evidence for 
$\overline{B}{}^0 \to D^{*0} \eta$ and $D^{*0}\omega$.  
The branching fractions are found to be 
${\cal B} (\overline{B}{}^0 \to D^0 \pi^0) = 
	(3.1 \pm 0.4 \pm 0.5) \times 10^{-4}$,
${\cal B} (\overline{B}{}^0 \to D^{*0} \pi^0) = 
	(2.7 \;^{+0.8}_{-0.7}\;^{+0.5}_{-0.6})\times 10^{-4}$,
${\cal B} (\overline{B}{}^0 \to D^0 \eta) = 
	(1.4\;^{+0.5}_{-0.4}\pm 0.3) \times 10^{-4}$,
${\cal B} (\overline{B}{}^0 \to D^0 \omega) = 
	(1.8 \pm 0.5 \;^{+0.4}_{-0.3}) \times 10^{-4}$, 
and we set 90\% confidence level upper limits of 
${\cal B} (\overline{B}{}^0 \to D^{*0} \eta) < 4.6\times 10^{-4}$ and
${\cal B} (\overline{B}{}^0 \to D^{*0} \omega) < 7.9 \times 10^{-4}$.
The analysis is based on a data sample of 21.3 fb$^{-1}$ collected 
at the $\Upsilon(4S)$ resonance by the Belle 
detector at the KEKB $e^{+} e^{-}$ collider. 
\vskip1pc \hspace*{-20pt}
\end{abstract}



\newpage
\pagestyle{plain}

\begin{figure}
\begin{center}
\begin{minipage}{4.1in}
\epsfig{file=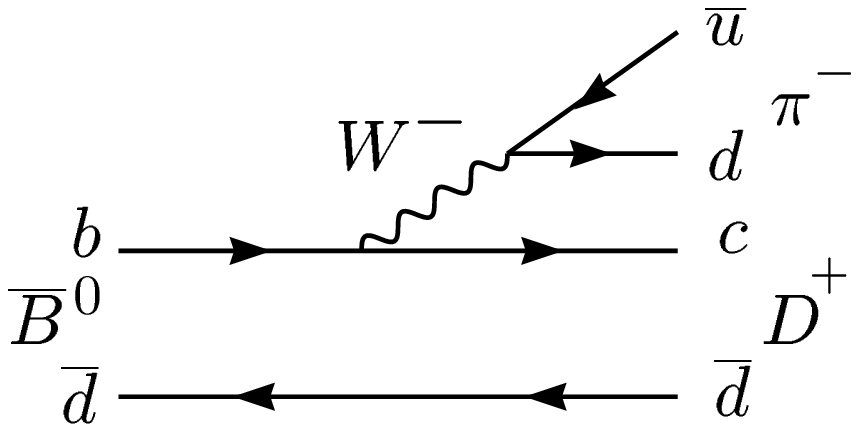,width=2in}
\epsfig{file=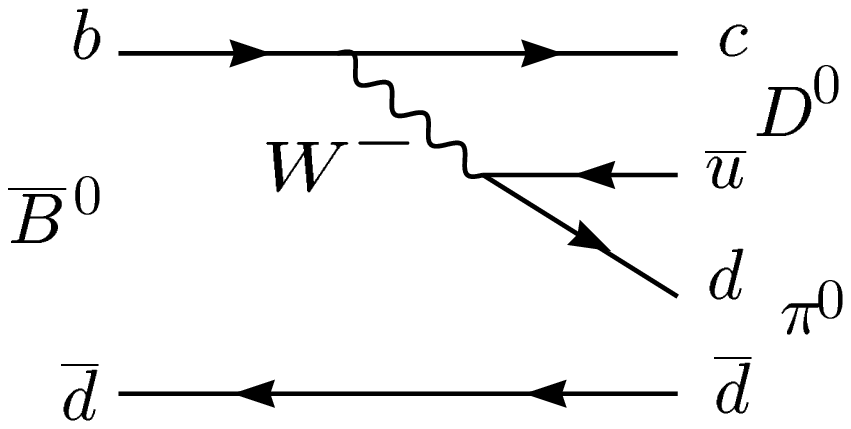,width=2in}
\end{minipage}
\medskip
\caption{The external (left) and internal (right) spectator 
diagrams for $\overline{B}\to D\pi$ decays.}
\label{dpi-feynman}
\end{center}
\end{figure}

Decay modes 
such as $\overline{B}{}^0 \to D^{(*)+}\pi^-$ were 
amongst the
first $B$ meson decays to be fully reconstructed. 
However, the $\overline{B}{}^0\to D^{(*)0} h^0$ decay modes, where $h^0$ 
represents a
light neutral meson, have not been observed to date. 
A recent search by the CLEO 
Collaboration\cite{cleo-prd57} yielded only upper limits.
As shown in Fig.~\ref{dpi-feynman}, these decays are expected to 
proceed via an internal spectator diagram and be suppressed relative to decays
proceeding via the external spectator diagram, since the color of the
$\overline{u}$ antiquark produced by the weak current
must complement the color of the $c$ quark.
The contribution of the $W$-exchange diagram is usually assumed to be
negligible~\cite{neubert}. 
Studies of such color-suppressed decay modes are useful for testing
models of hadronic $B$ meson decay and provide information on
final-state interactions.  
Final state interactions, if also present in $B$ decays to charmless 
modes, could have significant impact on $CP$ violating rate
asymmetries~\cite{Hou}. Furthermore, studies of $\overline{B}{}^0 \to
D^{(*)0} h^0$ where the $D^0$ decays to a CP eigenstate will provide
access to the $CP$ violation parameter $\sin{2\phi_1}$ in $b\to c
\overline{u}d$ processes~\cite{Dunietz}. Thus, measuring the 
strength of color-suppressed modes can have profound implications on $B$ 
physics and $CP$ violation.

In this paper, we report on a search for the color-suppressed
$\overline{B}{}^0 \to D^0h^0$ and $D^{*0}h^0$ \cite{ch-conj} decays, 
where the neutral meson $h^0$ is either a $\pi^0$, $\eta$, or $\omega$.
We make first observations of the modes $\overline{B}{}^0 \to D^0\pi^0$,
$D^{*0}\pi^0$, $D^0\eta$, and $D^0\omega$ 
and find evidence for the modes $\overline{B}{}^0\to D^{*0}\eta$ and 
$D^{*0}\omega$.
The data used in this analysis were
collected with the Belle detector\cite{NIM} at KEKB\cite{kekb},
a double storage ring that collides 8 GeV electrons with 3.5 GeV positrons 
with a 22 mrad crossing angle.
The data sample corresponds to an integrated luminosity of 
$21.3$ fb$^{-1}$ at the $\Upsilon(4S)$ resonance,
which contains 23.1 million $B\overline{B}$ pairs,
and $2.3$ fb$^{-1}$ taken 60 MeV below the resonance.


Belle is a general-purpose detector with a 1.5 T superconducting
solenoid magnet.  
Charged particle tracking, covering 92\% of the total center-of-mass (CM) 
solid angle, is provided by the Silicon Vertex Detector (SVD) consisting of
three concentric layers of double-sided silicon strip
detectors and a 50-layer Central Drift Chamber (CDC). 
Charged hadrons are distinguished by combining the responses
from an array of Silica Aerogel \v Cerenkov Counters (ACC), 
a Time of Flight Counter system (TOF), and $dE/dx$ measurements in the CDC. 
The combined response provides $K/\pi$ separation of at least 2.5$\sigma$
for laboratory momenta up to 3.5 GeV/$c$.  
Photons and electrons are detected in an array of 8736 
CsI(Tl) 
crystals (ECL) located inside the magnetic field and
covering the entire solid angle of the charged particle tracking system.  
The 1.5~T magnetic field is returned via an iron
yoke, instrumented to detect muons and $K_L$ mesons (KLM).
The KLM consists of alternating layers of resistive plate chambers and
4.7 cm thick steel plates.


We reconstruct light neutral mesons $h^0$ using the 
$\pi^0 \to \gamma \gamma$, $\eta \to \gamma\gamma$, $\eta\to \pi^+\pi^-\pi^0$ 
and $\omega\to\pi^+\pi^-\pi^0$ decay channels.
Charged tracks are required to have impact parameters that are
within $\pm$5 cm of the interaction point along the positron beam axis
and 1 cm in the transverse plane.
We reject tracks that are consistent with being electrons or muons.
The remaining tracks are identified as pions or kaons according to a 
kaon to pion likelihood ratio. 
Candidate $\pi^0$ mesons are reconstructed from pairs
of photons in the ECL if the $\gamma$ pairs 
have invariant mass inside a $\pm 3\sigma$ ($\sigma = 5.4~{\rm
MeV}/c^2$) mass window around the $\pi^0$ peak.
The $\pi^0$ daughter photons are required to have energies greater than
50 MeV. The $\pi^0$s are then  
constrained to the nominal $\pi^0$ mass \cite{pdg}.
Candidate $\eta$ mesons are required to
have invariant masses within $\pm 2.5\sigma$ of the $\eta$ peak, 
where $\sigma$ is $10.6~{\rm
MeV}/c^2$ and $3.4$ MeV/$c^2$ for the $\gamma\gamma$ and
$\pi^+\pi^-\pi^0$ modes, respectively. 
For the $\pi^+\pi^-\pi^0$ mode, 
the $\pi^+\pi^-$ pair is constrained to a vertex.
Both photons from the $\eta \to \gamma\gamma$ mode are 
required to have $E_{\gamma} > 100$ MeV and
the energy asymmetry of the daughter photons,
$\frac{|E_{\gamma_1}-E_{\gamma_2}|}{E_{\gamma_1}+E_{\gamma_2}}$,
is required to be less than 0.8.
We remove $\eta$ candidates if either of the daughter photons
can be combined with any other photon with 
$E_{\gamma} > 100$ MeV to form a $\pi^0$ candidate.
The $\eta$ candidates are further constrained to the known $\eta$
mass\cite{pdg}.
Candidate $\omega$ mesons are constructed from $\pi^+\pi^-\pi^0$ 
combinations where the $\pi^+ \pi^-$ pair must form
a vertex, and the CM momentum of the $\pi^0$ is required to be 
greater than 350 MeV/$c$ to reduce the large 
combinatorial background from low energy photons.
The invariant mass of the $\pi^+\pi^-\pi^0$ combination is required to be
within $\pm 30$ MeV/$c^2$ of the nominal $\omega$ mass~\cite{pdg}
(the natural width of the $\omega$ is $8.9$ MeV/$c^2$). 


We reconstruct $D^0$ mesons in the $D^0\to K^-\pi^+$, 
$K^-\pi^+\pi^0$, and $K^-\pi^+\pi^-\pi^+$ decay modes.
The CM momentum of the $\pi^0$ from $D^0\to K^-\pi^+\pi^0$ decay 
is required to be greater than 300 MeV/$c$. 
The invariant mass of the $D^0$ candidate is required to be within $\pm
2.5\sigma$ of the measured $D^0$ mass where $\sigma$, the
$D^0$ mass resolution, varies between 5.5 and 13 MeV$/c^2$
depending on the decay mode. 
A mass and vertex-constrained kinematic fit is then performed on the
$D^0$ candidates. 
$D^{*0}$ candidates are reconstructed in the $D^{*0}\to D^0\pi^0$ 
decay mode; 
the minimum photon energy requirement is reduced to 20 MeV for this case. 
The mass difference, $\delta m = M(D^0 \pi^0) - M(D^0)$, is 
required to be within $\pm 2.5\sigma$ ($\sigma = 0.82 {\rm~MeV}/c^2$) of 
its nominal value.


We combine $D^0$s or $D^{*0}$s with $h^0$ meson candidates to form 
$\overline{B}{}^0$ candidates. 
Two kinematic variables are used to identify
signal candidates, the beam-constrained mass   
$M_{bc} = \sqrt{(E_{\rm beam}^{\rm CM})^2 - (p_{B}^{\rm CM})^2}$
and the energy difference 
$\Delta E = E_{B}^{\rm CM} - E_{\rm beam}^{\rm CM}$, 
where $E_{B}^{\rm CM}$ and $p_{B}^{\rm CM}$ are the CM energy
and momentum of the $\overline{B}{}^0$ candidate, and 
$E_{\rm beam}^{\rm CM} = \sqrt{s}/2 = 5.29 {\rm~GeV}$.
The typical $M_{bc}$ resolution is 3 MeV$/c^2$; the $\Delta E$ 
resolution ranges from 17 to 35~MeV, depending on the decay mode.
When more than one $\overline{B}{}^0$ candidate is found in an event,
the candidate with the minimum $\chi^2$ is chosen, where 
$\chi^2 = \chi^2_{D^0} + \chi^2_{h^0} (+ \chi^2_{\delta m}$).
Here $\chi^2_{D^0}$ is from the kinematic
fit to the $D^0$, $\chi^2_{h^0}$ is from the kinematic
fit to the $h^0$ for the $\pi^0$ or the $\eta$, while 
$\chi^2_{h^0} = (\Delta(M_{\omega})/\sigma(M_{\omega}))^2$ 
for the $\omega$. Here $\Delta(M_{\omega})$ is the mass difference
between measured and nominal mass values and  
$\sigma(M_{\omega})$ is the measured resolution.  
For the $D^{*0}h^0$ modes, $\chi^2_{\delta m}$, 
defined as $(\Delta(\delta m)/\sigma({\delta m}))^2$,
is included in the best candidate selection. 


The background from continuum $e^+ e^- \to q\overline{q}$
production is suppressed in the following way.  
We form a Fisher discriminant\cite{fisher} containing seven variables 
that quantify event topology. 
The Fisher variables include the angle between the thrust axis~\cite{thrust} 
of the $B$ candidate and the thrust axis of the rest of the event 
($\cos{\theta_{T}}$), the $S_\perp$ variable\cite{sper}, 
and five modified\cite{etapk} Fox-Wolfram moments\cite{fw}.
In the $D^{*0} \pi^0$ and $D^0 \omega$ modes, 
helicity provides additional discrimination.
For the $D^{*0}\pi^0$ mode, we define ${\cal H}_{D^{*0}}$ 
as the cosine of the angle between the 
$D^{*0}$ flight direction and the direction of the $\pi^0$
in the $D^{*0}$ rest frame and require $|{\cal H}_{D^{*0}}| > 0.4$.
For the $D^0\omega$ mode, we define ${\cal H}_{\omega}$
as the cosine of the angle between the $B$ flight direction and 
the normal to the $\omega$ decay plane in the $\omega$ rest frame.
We also define a variable $\cal A$
as the absolute value of the
cross product of the two charged pion momentum vectors,
$| \vec{P}_{\pi^+} \times \vec{P}_{\pi^-} |$, 
in the $\omega$ rest frame.
To suppress the $q\overline{q}$ background, we take the 
$B$ flight direction and the Fisher discriminant, and 
for modes with $\omega$ mesons we include 
the variable $\cal A$.
For the $D^0\omega$ mode, we also include the variable ${\cal H}_{\omega}$. 
These variables are then combined to form signal (S) and background
(BG) probability density functions (PDFs).
Signal PDFs are determined using Monte Carlo (MC), and background PDFs
are obtained from $M_{bc}$ sideband data. 
The PDFs are multiplied to form a signal (background)
likelihood ${\cal L}_{\rm S (BG)}$, and a selection is applied on
the likelihood ratio 
${\cal L}_{\rm S}/({\cal L}_{\rm S}+{\cal L}_{\rm BG})$. 
This requirement, which has a typical efficiency of 70\%,
removes more than 90\% of the $q\overline{q}$ background.


In addition to the $q\overline{q}$ background, there are large
background contributions from color-favored $B \to D^{(*)} (n\pi)^-$
decays and feed-down from $D^{*0} h^0$ to $D^0 h^0$ modes.
The color-favored $B \to D^{(*)} (n\pi)^-$ events give rise to 
two kinds of backgrounds: those with the same final state particles as
signal events, and those with missing or extra particles.
The former mimics the signal distributions in both $M_{bc}$ and $\Delta E$,
and selection cuts on variables other than these (discussed below)
are needed to suppress these backgrounds.
The backgrounds with different final state particles arise 
when a pion is missed or a spurious pion is added. 
In this case, the background events may not be distinguishable
in the $M_{bc}$ distribution if the missing or extra
pion has very low momentum, but the $\Delta E$
distribution provides a useful discriminant because
of the missing pion rest-mass energy.

The $\overline{B}{}^0 \to D^{*+}\rho^-$ mode can give the same 
final states as $D^0\omega$ and $D^0 \eta$. 
However, more than 99\% of the background events are removed by the
$\omega$ and $\eta$ mass cuts since the invariant mass of the three pions
rarely falls within the $\omega$ or $\eta$ mass region.

The $B^-\to D^{(*)0}\rho^-$ final state can contaminate
the $D^{(*)0}\pi^0$ mode if the $\rho^-$ decay produces a fast $\pi^0$.
The background events from $D^{0}\rho^-$ to $D^0\pi^0$ ($D^{*0}\rho^-$
to $D^{*0}\pi^0$) have $\Delta E$ values that are below the signal
region because of the slow $\pi^-$ that is missed in the reconstruction.
However, the signal contamination is not negligible
due to the large branching fraction of $B^-\to D^{(*)0}\rho^-$.
The $D^{0}\rho^-$ mode can also cause a background to $D^{*0}\pi^0$
if a low momentum $\pi^0$ is used to form a $D^{*0}$ candidate. 
This case is similar to the case where the final state particles are
the same, and causes a few events in the signal region.
About half of these events are removed by rejecting events that can be
reconstructed as $B^- \to D^{(*)0}\rho^-$.
This reduces the systematic error when fitting for the signal
yields and only removes a few percent of signal events.

The $B^-\to D^{(*)0}\rho^-$ mode can also contaminate
the $D^{(*)0}\eta$ channel if a photon from the fast $\pi^0$ 
is combined with another photon to form an $\eta$ candidate.
The contributions of these backgrounds in the $\eta$ channel, as
well as the feed-across from the $D^{(*)0}\pi^0$ mode,
are minimized by the $\pi^0$ veto described earlier.

We also check the background contributions from 
$B^-\to D^{(*)0}\rho^{\prime -}$ ($\rho^{\prime -}\to
\omega\pi^-$) decays that have
been reported recently by the CLEO Collaboration\cite{cleo0103021}.
This two-body decay produces high momentum $D^{(*)0}$s and $\omega$s
that may fake signal events. Monte Carlo studies indicate that the
remaining background events are shifted in $\Delta E$ by more 
than the mass of the missing pion and thus are distinguished from
signal events by the fit to the $\Delta E$ distribution.
The $D^{0}\rho^{\prime -}$ mode can also contaminate the $D^{*0}\omega$ mode
if the $\pi^-$ from $\rho^{\prime -}$ is replaced by a $\pi^0$.
However, since the $\pi^-$ from $\rho^{\prime -}$ decay carries 
sizable momentum, the kinematics of the final state particles are
different from that of the $D^{*0}\omega$ signal, and the expected
background is small. 

\begin{figure}
\begin{center}
\vspace{-10pt}
\epsfig{file=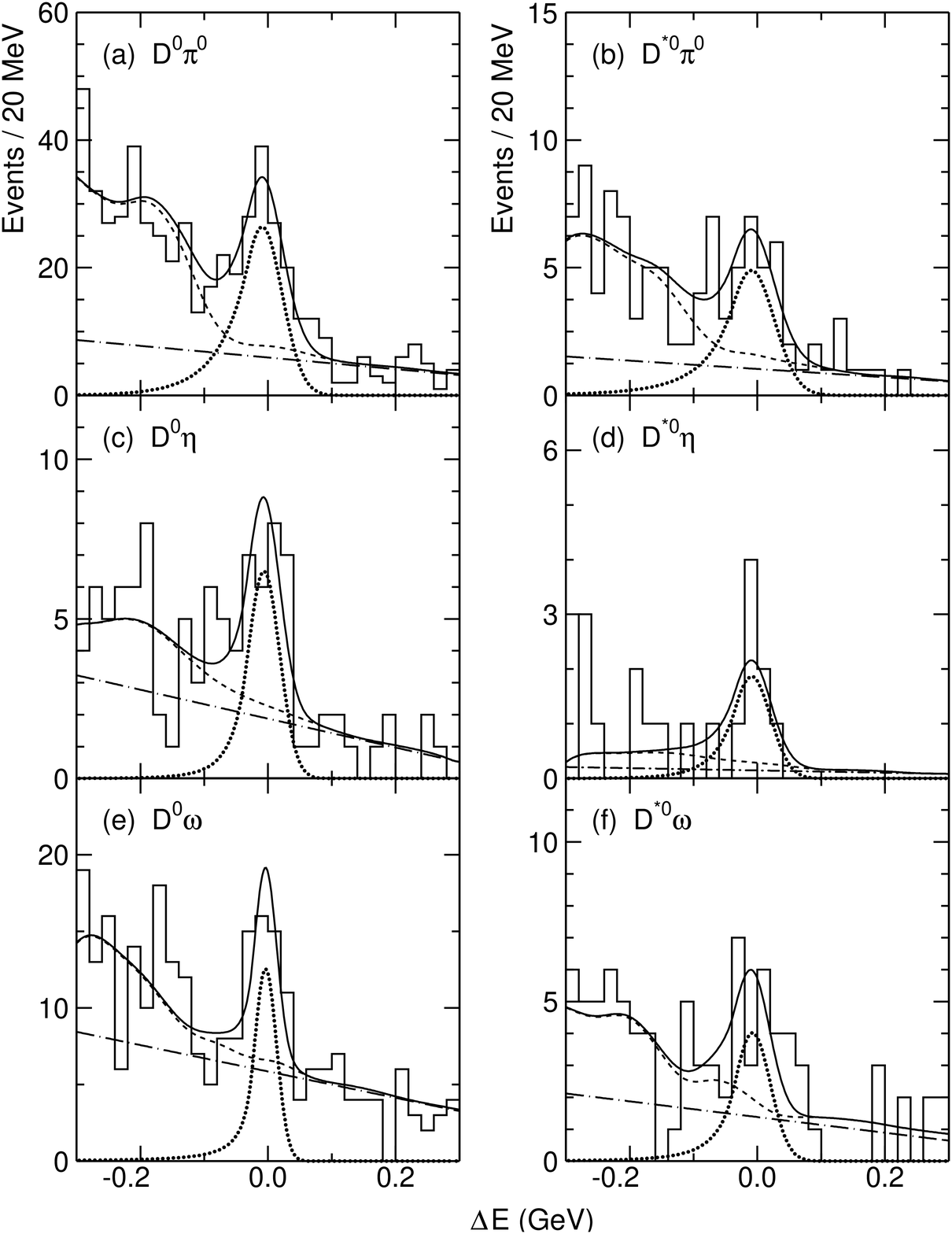, width=4in}
\medskip
\caption{The $\Delta E$ distributions for (a) $D^{0}\pi^0$, 
(b) $D^{*0}\pi^0$, (c) $D^{0}\eta$, 
(d) $D^{*0}\eta$, (e) $D^{0}\omega$, and 
(f) $D^{*0}\omega$.
The solid lines show the fit results, the dotted lines are the signals, 
and the dashed lines show the
sum of the feed-across and the combinatorial background,
with the latter shown separately as the dash-dotted lines.}
\label{fig-d0h0}
\end{center}
\end{figure}

The $\Delta E$ distributions for the various $D^{(*)0}h^0$ decays,
after the application of all selection requirements and 
with $M_{bc}$ between 5.27 GeV/$c^2$ and 5.29 GeV/$c^2$,
are shown in Fig.~\ref{fig-d0h0}.
The plots are fitted to the signal and background functions
using a binned maximum likelihood.
The signal shape is an empirically determined
parameterization~\cite{cbline} with parameters obtained via MC.
We observe good agreement between data and MC for the $\Delta E$
distributions of color-favored decays such as $D^{*+}\rho^-$ and
$D^0\rho^-$.
The background functions include a combinatorial component and a
color-favored component. 
For the $D^0 h^0$ modes, we also include a component for feed-down from 
color-suppressed $D^{*0} h^0$ modes.
The contribution from $D^0 h^0$ to $D^{*0} h^0$ modes is found to
be negligible.
The combinatorial component is taken to be a first-order polynomial 
with a slope determined from the $\Delta E$ shape of the $M_{bc}$
sideband (5.20 GeV/$c^2 < M_{bc} < 5.26$ GeV/$c^2$) data. 
The shapes of the color-favored and feed-down components are modeled by MC.
The area of the feed-down component from $D^{*0}h^0$ to
$D^0h^0$ is fixed by the obtained signal yield from $D^{*0}h^0$ with
the estimated feed-down efficiency in MC.
For the $D^{(*)0}\omega$ mode, the $D^{(*)0}\rho^{\prime -}$ components
are also fixed by MC using the measured branching
fractions\cite{cleo0103021}. 
The normalizations of the signal and background components
are free parameters.

Table~I lists the signal yield, statistical significance, reconstruction
efficiency, and branching fraction for each $D^{(*)0}h^0$ mode.  In
addition, the estimated number of events in the region of -0.1 GeV$<\Delta
E<$ 0.1 GeV due to backgrounds from generic $B\overline{B}$ decays, 
$D^{*0} h^0$ feed-down, and $q\overline{q}$ production are given.
The systematic errors due to fitting are obtained 
by varying the parameters of the fitting functions within $1\sigma$ of
their nominal values. The change in the signal yield from each variation is
added in quadrature to obtain systematic errors from the fit.  
These are between 5\% and 15\% depending on the decay mode. 
The statistical significance is defined as 
$\sqrt{-2{\rm ln}({\cal L}(0)/{\cal L}_{\rm max})}$
where ${\cal L}_{\rm max}$ is the likelihood at the nominal signal
yield and ${\cal L}(0)$ is the likelihood with the signal yield fixed
to zero.  We observe signals for $\overline{B}{}^0 \to D^0 \pi^0$, 
$D^{*0}\pi^0$, $D^0 \eta$, and $D^0 \omega$ decays with 
more than $4\sigma$ significance.
Independent fits to the $M_{bc}$ distributions, after
subtracting $B\overline{B}$ background, confirm these results. 
We find evidence for $\overline{B}{}^0 \to D^{*0} \eta$ and $D^{*0} \omega$ 
with more than $3\sigma$ significance.  
For decay modes with significance less than 4$\sigma$, we give 90\%
confidence level upper limits (UL) on the signal yields ($N_{\rm
S}^{\rm UL}$) from the relation 
$\int_0^{N_{\rm S}^{\rm UL}}{\cal L}(N_{\rm S}) \, dN_{\rm S} \,/\, 
\int_0^{\infty}{\cal L}(N_{\rm S}) \, dN_{\rm S} =0.9$, 
where ${\cal L}(N_{\rm S})$ denotes the maximum
likelihood with the signal yield fixed at $N_{\rm S}$. 

The efficiencies for each mode are obtained by Monte Carlo and 
calibrated by a detailed study of detector performance.
In particular, studies of tracking, $\pi^0$ detection, and particle
identification give systematic errors in the 
detection efficiencies of low momentum $\pi^0$, energetic $\pi^0$,
$\eta$, and $\omega$ mesons of 10.7\%, 7.3\%, 9.6\%, and 9.5\% respectively. 
The systematic error on the $D^0$ meson reconstruction efficiency is 
estimated to be 12.7\% 
by comparing the observed yield of $B^-\to D^0\pi^-$ events 
with the expected yield using known branching fractions~\cite{pdg}.
A 5\% systematic error on the likelihood ratio requirement is determined
by applying the same procedure to the $B^- \to D^{*0}\pi^-$
sample and comparing the effects on data and MC.
The final systematic errors of the branching
fractions include the errors in fitting,
reconstruction efficiency, cut efficiency for background suppression,
and the number of $B\overline{B}$ pairs. 
Assuming the number of $B^0\overline{B}{}^0$ and $B^+B^-$ pairs to be equal,
we calculate the branching fractions for various decay modes given in 
Table~\ref{yields-table}I.  
The branching fraction upper limits are calculated by increasing
$N_{\rm S}^{\rm UL}$ and reducing the efficiency by their systematic errors.


Our measurement of ${\cal B}(\overline{B}{}^0 \to D^0\pi^0)
 = (3.1 \pm 0.4 \pm 0.5) \times 10^{-4}$
is above the previous UL of $1.2 \times 10^{-4}$ 
obtained by CLEO\cite{cleo-prd57}.
It is also considerably higher than the 
factorization prediction of $0.7 \times 10^{-4}$\cite{neubert}.
This could indicate the presence of final state interactions,
or other corrections to factorization.
It is customary to decompose\cite{neubert} the
$B^- \to D^0\pi^-$ and $\overline{B}{}^0 \to D^+\pi^-$, $D^0\pi^0$ decay
amplitudes into isospin $\frac{1}{2}$ and $\frac{3}{2}$ components.
Our measurement of ${\cal B}(\overline{B}{}^0 \to D^0\pi^0)$,
together with the known branching fractions~\cite{pdg} of the other two modes,
suggest a rescattering phase difference between 
isospin $\frac{1}{2}$ and $\frac{3}{2}$ amplitudes that is $31^\circ \pm 7^\circ$. 
A similar value, $32^\circ \pm 8^\circ$, is obtained for the $\overline{B} \to D^{*}\pi$ system.

Our measurements of 
${\cal B}(\overline{B}{}^0\to D^0\eta) = 
(1.4\;^{+0.5}_{-0.4}\pm 0.3)\times 10^{-4}$ and
${\cal B}(\overline{B}{}^0\to D^0\omega) = 
(1.8 \pm 0.5\;^{+0.4}_{-0.3})\times 10^{-4}$ 
are also higher than 
the factorization predictions of $0.5 \times 10^{-4}$ and 
$0.7 \times 10^{-4}$\cite{neubert}.
The central values of the two less significant modes show the same pattern.
The results for these modes cannot be accommodated by the
aforementioned elastic rescattering phase.

In summary, using 23.1 million $B\overline{B}$ events collected with the
Belle detector, we report the first observations of color-suppressed 
$\overline{B}{}^0 \to D^0 \pi^0$, $D^{*0} \pi^0$, $D^{0} \eta$, 
and $D^0\omega$ decays,
and evidence for $\overline{B}{}^0 \to D^{*0} \eta$ and $D^{*0}\omega$ modes. 
All these modes have similar branching fractions with
central values between 1.4 $\times 10^{-4}$ and 3.1 $\times 10^{-4}$,
as given in Table~\ref{yields-table}I.
They are all consistently higher than recent theoretical 
predictions based on the factorization hypothesis.  
This may be accounted for by additional corrections to the
factorization models, or by non-factorizable effects such as final
state interactions.


We wish to thank the KEKB accelerator group for the excellent
operation of the KEKB accelerator. We acknowledge support from the
Ministry of Education, Culture, Sports, Science, and Technology of Japan
and the Japan Society for the Promotion of Science; the Australian
Research Council and the Australian Department of Industry, Science and
Resources; the Department of Science and Technology of India; the BK21
program of the Ministry of Education of Korea and the Center for the
High Energy Physics sponsored by the KOSEF; the Polish
State Committee for Scientific Research under contract No.2P03B 17017;
the Ministry of Science and Technology of Russian Federation; the
National Science Council and the Ministry of Education of Taiwan; and
the U.S. Department of Energy.

\newpage
\begin{table}
\label{yields-table}


\caption{The obtained signal yields, statistical significances ($\Sigma$),
background events from generic $B\overline{B}$ decays ($B\overline{B}$ bg), 
$D^{*0}h^0$ feed-down ($D^{*0}h^0$), and $q\overline{q}$ production ($q\overline{q}$ bg),
reconstruction efficiencies ($\epsilon$) 
(including the sub-decay branching fractions),
branching fractions ($\mathcal{B}$), and $90\%$ confidence level upper 
limits (UL) for $\overline{B}{}^0\to D^{*0}\eta$ and $D^{*0}\omega$ 
are listed together with the theoretical predictions (Th)~[2].
The background events are estimated under the signal peak 
(in the region of -0.1 GeV$< \Delta E <$ 0.1 GeV).}
\medskip
\begin{tabular}{llccccclcc}
Mode & Signal Yield & $\Sigma$ & $B\overline{B}$ bg 
& $D^{*0}h^0$ & $q\overline{q}$ bg
& $\epsilon$(\%) & $\mathcal{B}$ ($\times 10^{-4}$) & UL ($\times 10^{-4}$) 
& Th ($\times 10^{-4}$) \\ \hline
$D^0\pi^0$      & $126.2\;^{+16.1}_{-15.5}\;^{+7.2}_{-5.2}$ 
	& 9.3 & 26.7 & 1.3 & 145.6 & 1.79 
	& $3.1\pm 0.4\pm 0.5$  & -- & 0.7\\
$D^{*0}\pi^0$   & $26.4\;^{+7.7}_{-7.1}{}\;^{+1.6}_{-2.2}$ 
	& 4.1 &  5.9 & --  &  10.4 & 0.42 
	& $2.7\;^{+0.8}_{-0.7}\;^{+0.5}_{-0.6}$ 
	& -- & 1.0 \\	\hline
$D^0\eta^{*}$       & $ 22.1\;^{+7.0}_{-6.3}\;^{+2.0}_{-1.8}$ 
        & 4.2 & 3.4 & 0.7 & 19.1 & 0.67 
	& $ 1.4\;^{+0.5}_{-0.4} \pm 0.3 $ 
	& -- & 0.5\\
$D^{*0}\eta$    & $ 7.8\;^{+3.6}_{-3.0} \pm 0.7 $ 
        & 3.3 & 1.4 & --  & 1.5  & 0.17 
	& $ 2.0\;^{+0.9}_{-0.8} \pm 0.4 $ 
	& 4.6 & 0.6 \\ 	\hline
$D^0\omega$     & $32.5\;^{+9.4}_{-8.6}\;^{+4.0}_{-3.1}$ 
	& 4.4 & 5.3(2.3)$^{\dag}$ & 1.4 & 58.5 & 0.80 
	& $1.8\pm 0.5\;^{+0.4}_{-0.3} $ 
	& -- & 0.7\\
$D^{*0}\omega$  & $16.1\;^{+6.8}_{-6.0}\pm 2.4$ 
	& 3.0 & 5.3(1.5)$^{\dag}$ & --  & 13.8  & 0.23 
	& $3.1\;^{+1.3}_{-1.1}\pm 0.8 $ 
	& 7.9 & 1.7\\
\end{tabular}
{\footnotesize ${}^{*}$ For decay modes with $\eta$ mesons, about 70\%
of the signal is reconstructed in the $\gamma \gamma$ channel
and the remainder is in the $\pi^+\pi^-\pi^0$ channel.}
{\footnotesize $^{\dag}$ The values in parentheses are the estimated
background from $B^-\to D^{(*)0}\rho'^-$ decays.}
\vskip1pc
\end{table}

\begin{thebibliography}{99}
\bibitem{ch-conj}{
	Charge conjugate modes are implicitly included throughout the
	paper.
}
\end{thebibliography}
\end{document}